# Structural, magnetic and superconducting characterization of the CuNi/Nb bilayers of the S/F type using Polarized Neutron Reflectometry and complementary techniques

Yu. Khaydukov[1,3*], R. Morari[2,5], L. Mustafa[1], J.-H. Kim[1], T. Keller[1,3], S. Belevski[2], A. Csik[4], L. Tagirov[5], G. Logvenov[1], A. Sidorenko[2], B. Keimer[1]

[1]*Max-Planck-Institut für Festkörperforschung, D 70569 Stuttgart, Germany*

[2]*Institute of Electronic Engineering and Nanotechnologies ASM, MD2028, Kishinev, Moldova*

[3]*Max Planck Society Outstation at FRM-II, D-85747 Garching, Germany*

[4]*Institute for Nuclear Research, Hungarian Academy of Sciences, Debrecen, Hungary*

[5]*Solid State Physics Department, Kazan Federal University, 420008 Kazan, Russia*

*E-mail: y.khaydukov@fkf.mpg.de

**Abstract**

Structural, magnetic, and superconducting properties of S/F bilayers Nb/$Cu_{40}Ni_{60}$ deposited on silicon substrate have been characterized using Polarized Neutron Reflectometry and complementary techniques. The study allowed to determine real thicknesses of the S and F layers as well as the r.m.s. roughness of the S/F interfaces. The latter does not exceed 1 nm, showing the high quality of the S/F interface. Using SQUID and a mutual inductance setup we determined the superconducting transition temperatures of the samples, which are in agreement with the literature data. Using of PNR for the single S layer allowed to determine the screening length $\lambda$ of the superconducting layer, $\lambda = 120$ nm. This value is higher than the London penetration depth for pure niobium which may indicate that the superconductor is in the dirty limit. PNR and SQUID studies of magnetic properties of the CuNi layer have shown the presence of ferromagnetism in all investigated samples.

## 1. Introduction

A proximity of a superconductor (S) and a ferromagnet (F) leads to appearance of a great number of intriguing phenomena, such as spatial oscillation of electron density of states, π type S/F/S Josephson junctions, F/S/F spin valves, etc. (see reviews [1,2]). The interaction between superconducting and ferromagnetic order parameters at S/F interface leads to the modification of both superconducting and magnetic properties of such systems. For example, it was shown in [3,4] that the superconducting transition temperature $T_c$ of a superconducting layer is an oscillating function of the ferromagnetic layer thickness $d_F$, which is caused by the finite momentum pairing of conduction electrons in a ferromagnetic layer. There is also an inverse effect of the magnetic order penetration into a superconductor (the so-called inverse proximity effect [5-7]). In this case, the magnitude and sign of the induced magnetization also depend on the ferromagnetic layer thickness and the quality of the superconductor/ferromagnetic interface.

In this work we have used Polarized Neutron Reflectometry (PNR) and complementary techniques (X-ray reflectometry (XRR), Secondary Neutral Mass Spectrometry (SNMS), SQUID magnetometry, mutual inductance setup) to study structural, magnetic, and superconducting properties of CuNi/Nb bilayers of the S/F type. In the PNR experiment reflectivities $R^{\pm}(Q)$ with spins along (sign "+") and opposite (sign "-") to a direction of the external magnetic field are measured as a function of the momentum transfer $Q$. The reflectivities can be expressed as $R^{\pm}(Q) \sim |\int [4\pi\rho(z) \pm cB(z)]\exp(iQz)dz|^2$, where $\rho(z)$ is the depth profile of the nuclear scattering length density (SLD), $B(z)$ is the depth-profile of the magnetic induction, $c = 2.9 \times 10^{-4}$ kGs nm$^2$ is the scaling factor. Since nuclear SLD is usually much greater than $cB(z)$, a special procedure is required to extract the magnetic signal. One approach is analysis of the so-called spin asymmetry $S \equiv [R^+ - R^-]/[R^+ + R^-]$, which is easily shown to be proportional to the Fourier transform of $B(z)$ [8]. Thus, PNR allows to extract both nuclear and magnetic depth profiles, and then, combined with complementary techniques to obtain comprehensive information about structural and magnetic properties of the S/F system.

## 2. Sample preparation

Samples were prepared in the magnetron sputtering machine Leybold Z-400 (with a residual pressure in the chamber of about $1.5 \times 10^{-6}$ mbar) on silicon (Si) wafers with (111) crystalline orientation. In total three targets were used: the pure niobium (99.99%) as a superconducting material, copper-nickel alloy (60% Ni -40% Cu) as a diluted ferromagnet, and the pure Si (99.99%) for growth of buffer and protective layers.

Sputtering was done in Argon (purity 99.999%, "Messer Griesheim") atmosphere of $8\times10^{-3}$ mbar pressure. The design of the deposition machine allows growing of the whole structure in one cycle without interruption of the vacuum, and this feature provides the structures of high quality with clean interfaces between the layers.

In order to remove contaminations, such as absorbed gases and oxides, the targets and substrates were pre-sputtered for 3-5 minutes before deposition of the S/F-structures. The deposition rate of the layers was: 4.5 nm/s for Nb, 3.5 nm/s for CuNi alloy, and 1.0 nm/s for Si. Part of the samples was prepared on two substrates simultaneously: Si ($20\times20$ mm$^2$) for neutron and X-ray reflectometry measurements and Si ($5\times5$ mm$^2$) for SQUID and superconducting critical temperature measurements. For obtaining homogeneous thickness of the layers on every substrate the target was wobbling during sputtering.

The stack sequence was as follows: first a thin film of amorphous silicon was sputtered on the top of the substrate in order to achieve a homogenous and flat surface. Second the layer of Nb was deposited followed by the deposition of the CuNi layer. As a final step, a cap silicon layer was deposited on top of the S/F heterostructure to protect it against oxidation. The parameters of structures are collected in Table 1.

## 3. Structure characterizations

Structural properties were investigated using X-ray and neutron reflectometry. Sample 5 without ferromagnetic layer has been additionally measured by SNMS to reveal the concentration profile of niobium atoms within the S layer. The obtained data allowed making a conclusion about high purity of the niobium layer. Neutron and X-ray measurements were conducted on the combined neutron/X-ray reflectometer NREX situated on the research neutron reactor FRM II (Garching, Germany). Part of the X-ray measurements was repeated on the D8 Bruker diffractometer. For both cases Cu-K$_\alpha$ line with the wavelength 1.54 Å was used. Neutron reflectivities were measured using the monochromatic (wavelength 4.3 Å) and polarized beam with divergence of the incident beam 0.02°. A closed cycle cryostat was used in the PNR experiment to cool down the sample to the temperatures around the superconducting transition temperature of Nb layers, $T_c$.

X-ray and neutron specular reflectivity curves for the sample No 2 are shown in Fig. 1. Both curves are characterized by the presence of a reflection plateau at low angles, and Kiessig oscillations caused by the interference on different interfaces inside the structure.

Fitting of the experimental XRR curve measured on sample No 2 using the model SLD depth profile (see inset in Fig. 1a) allows us to obtain thicknesses of the buffer layers (approximately 15 nm), r.m.s. roughness of the surface ($\sigma_1 = 0.4$ nm), the CuNi/Si interface ($\sigma_2 = 0.6$ nm) and the Si/Nb interface ($\sigma_4 = 0.5$ nm). Since the X-ray SLD of Nb ($6.38\times10^{-5}$ Å$^{-2}$) and CuNi ($6.44\times10^{-5}$ Å$^{-2}$) are very close (~1% of difference), XRR is sensitive to the total thickness of CuNi and Nb layers.

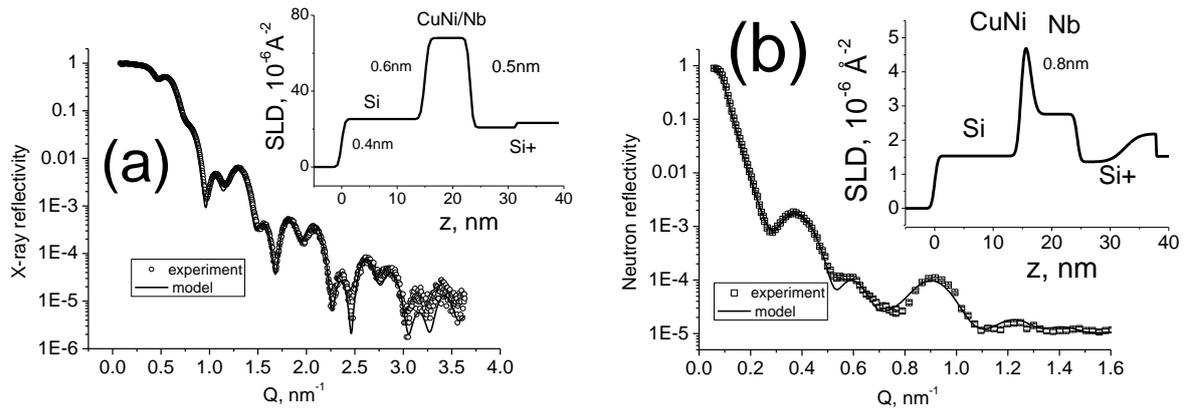

Fig. 1. X-ray (a) and neutron (b) experimental reflectivity curves (symbols) with the best-fit model curves (solid lines) for sample No 2. Corresponding SLD depth profiles are shown in the insets. The r.m.s. roughness for the every interface is shown by numbers in the vicinity of the corresponding interface.

After the fit of XRR, the neutron reflectivities were fitted. To fit the experimental curve by the model we have varied thicknesses and SLD of the CuNi and Nb layers as well as the r.m.s. height of the roughness of the S/F interface. All the remaining parameters were taken from the fit of the XRR data and kept fixed during the fit of the neutron reflectivities. The thicknesses of the CuNi layer $d_F$ and Nb layer $d_S$ were constrained to have the total thickness obtained from XRR. The fit allowed us to determine thicknesses of S ($d_S = 8.3\pm0.8$ nm) and F layers ($d_F = 3.5\pm0.1$ nm) and the r.m.s. roughness on the S/F interface, $\sigma_3 = 0.8$ nm$\pm0.3$ nm. All other samples have been fitted in a similar way. The data treatment has shown that in all measured samples the r.m.s. roughness of the S/F interface do not exceed 1 nm, showing high quality of the S/F interface (see Table 1). The corresponding interfaces with the Si buffer and cap layers can be rougher, reaching a value of 3 nm.

## 4. Superconducting properties

The SQUID magnetometry and mutual inductance setup were used to determine the superconducting transition temperature $T_c$. The obtained temperatures are in agreement with the values given in literature for the corresponding $d_S$ and $d_F$ [4].

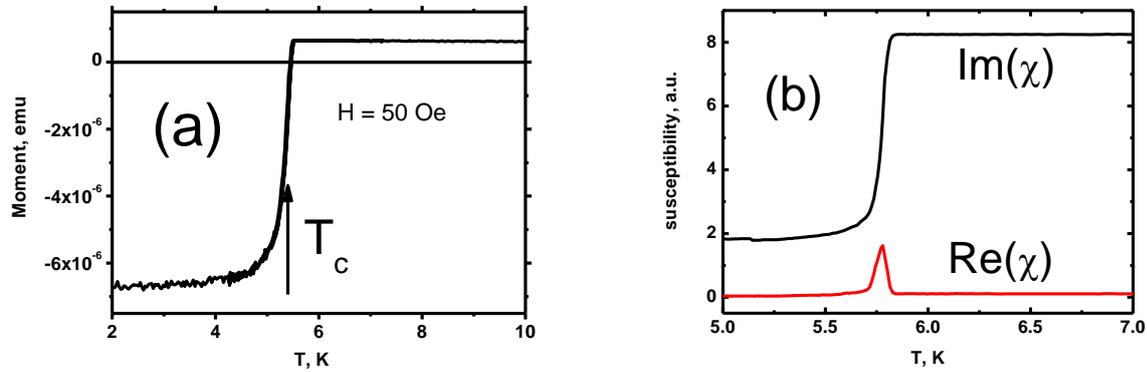

Fig. 2. (a) The temperature dependence of the field cooled SQUID magnetic moment of sample 2 with the in-plane external field $H = 50$ Oe. (b) Temperature dependence of the amplitude of the mutual inductance of sample 4.

PNR has been used to measure the penetration depth of the magnetic field into the S layer. For this study we have chosen sample 5 with relatively thick S layer having no ferromagnetic layer. The reflectivity curves measured above $T_c$ revealed no spin asymmetry, showing the absence of the magnetic signal. These curves were used to reconstruct the nuclear SLD profile. At $T = 3.5$K $= 0.43 T_c$ in magnetic field of $H = 4$ kOe, applied parallel to the sample surface, the spin asymmetry with the maximum amplitude $S_{max} = -5 \pm 1\%$ has been observed around the critical edge (inset in Fig. 3a).

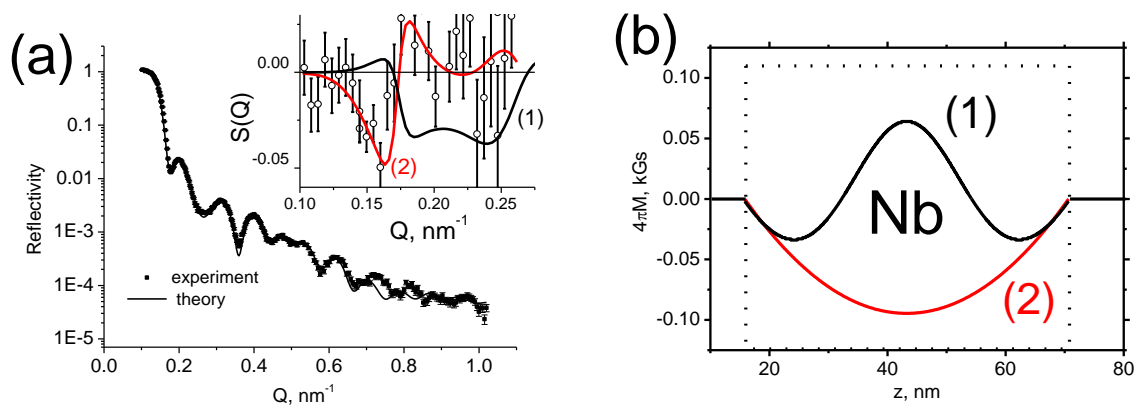

Fig. 3. (a) Experimental and model $R^+$ reflectivities measured on sample 5 at the temperature $T = 10$ K in magnetic field $H = 4$ kOe. Inset: Experimental (symbols) spin asymmetry measured on sample 5 in the magnetic field of $H = 4$ kOe at $T = 3.5$K. Solid lines (1) and (2) are the model spin asymmetries for the Abrikosov and Meissner states correspondingly (see Fig. 3b). (b) The depth profiles of the magnetization in the Abrikosov state (1) and the Meissner state (2) with $\lambda = 120$ nm.

Since the applied magnetic field was pretty high, we first tried to fit the experimental data assuming that the S layer is in the Abrikosov mixed state. A model with the single raw of the vortices with the core of 10nm in the center of the S film was assumed (see magnetic profile (1) in Fig. 3b). However, this model does not describe the data. Second, we tried to fit the experimental spin asymmetry by magnetic depth profile corresponding to the Meissner response of a thin superconducting film [9]

$$4\pi M(z) = H\left(\frac{\cosh\left(\frac{z-z_0-d_S/2}{\lambda}\right)}{\cosh\left(\frac{d_S}{2\lambda}\right)} - 1\right), \text{ where} \qquad (1)$$

$z_0$ is the coordinate of the top of the S layer, and $\lambda$ is the experimental screening length of the external magnetic field $H$. A good agreement between the experiment and the model is observed at $\lambda = 120 \pm 24$ nm. This value agrees within the error bar for the $\lambda$ parameter obtained in Ref. [10] for Nb(50nm) film using microwave surface impedance and with $\lambda = 110$ nm obtained by PNR on Nb(600nm) sample [11]. The absence of vortices in the S film at high magnetic field in our case is explained by small thickness comparing to the vortex size [12,13].

We also notice that the obtained screening length $\lambda = 120$ nm is 3 times higher than the London penetration depth for the bulk Niobium $\lambda_L = 47$ nm. A screening length which exceeds London penetration depth was already observed in Nb films in several PNR works, e.g. in [14,11]. In [11] this was explained by the fact that the experimentally measured value is $\lambda = \lambda_L(1 + \xi_0/l_S)^{1/2}$ where $\xi_0$ and $l_S$ are the superconducting coherence length and the electron mean free-path in the S layer, correspondingly. In the case of dirty superconductors ($\xi_0 \gg l_S$), the screening depth exceeds $\lambda_L$. For our case, using $\xi_0 = 42$ nm for bulk Nb, we obtain $l_S = 6.4$ nm.

It is also worth to comment that for relatively thick $d_S = 57$ nm S layer, the diamagnetic response is found to be only 2.5% of the applied external field (Fig. 3b). For the S/F samples with much thinner $d_S \approx 8$ nm the diamagnetic response is expected to be only 0.05%. For example, at the applied $H = 4$ kOe the diamagnetic response would be of order of 2 Gs which is below detection limit of PNR.

## 5. Magnetic properties

Magnetic properties of the samples were characterized by Quantum Design MPMS SQUID VSM magnetometer and by PNR. The magnetic hysteresis loop of sample 2 measured at $T = 15$ K with magnetic field applied parallel to the sample surface is presented in Fig. 4a.

The loop contains ferromagnetic signal from the CuNi layer and diamagnetic background from the Si substrate. The latter can be subtracted similar to [4]. As it can be seen from Fig. 4a, the saturation magnetic field is of the order of $H_{sat} \approx 1$ kOe. The remanent magnetic moment of the F layer, $m_{rem} = 6.4 \times 10^{-7}$ emu, is only 30% of the saturation magnetic moment $m_{sat} = 2.2 \times 10^{-6}$ emu. This tells us about the presence of perpendicular magnetic anisotropy in CuNi layer [15].

Spin asymmetry of the neutron reflectivity gives us another evidence of the presence of a ferromagnetic signal in thin CuNi films. The spin polarized neutron reflectivities $R^+$ and $R^-$ were measured at several temperatures above $T_c$ in the magnetic field $H = 4$ kOe, which is well above saturation. To increase the statistical accuracy we have summed up curves for all temperatures above $T_c$. This allows us to see clearly the spin asymmetry of the order of 2% with statistical accuracy of 0.4% (Fig. 4b). Such a small signal, according to the calculations, corresponds to the magnetization of $4\pi M = 0.2$ kGs. Best to our knowledge this is one of the smallest magnetization measured by PNR on a single ferromagnetic film with the thickness of less than 10 nm. Compare for example with Ref. [16], where magnetization $4\pi M = 0.7$ kGs was measured by PNR in the 1-nm-thick $Au_{97}Fe_3$ film.

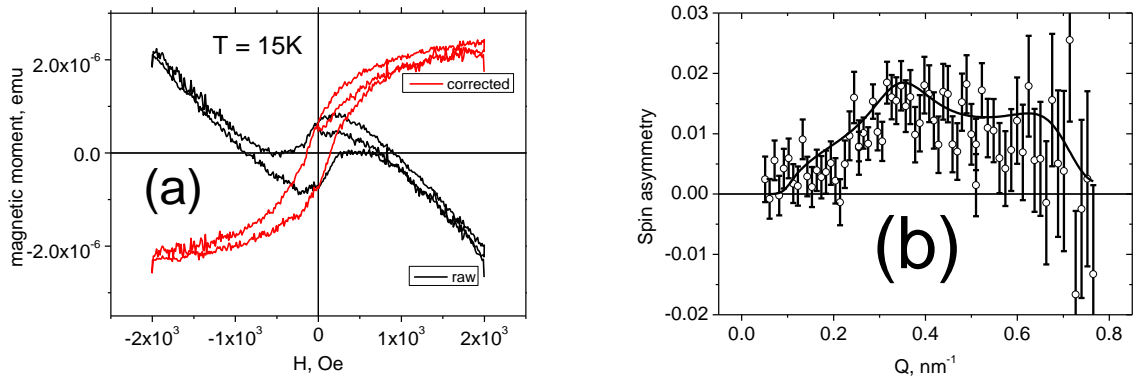

Fig. 4. (a) SQUID magnetization hysteresis loop of sample No 2 at 15 K. The magnetic field was applied in-plane. (b) Experimental spin asymmetry measured at $T > T_c$, $H = 4$ kOe on sample 3.

## 6. Summary

Structural, magnetic and superconducting properties of the S/F bilayers Nb($d_S$)/Cu$_{40}$Ni$_{60}$($d_F$) deposited on silicon substrate have been characterized using the Polarized Neutron Reflectometry and complementary techniques. In particular, comprehensive analysis of the neutron and X-ray reflectivities allowed extracting the thicknesses of the S and F layers as well as the r.m.s. roughness of the S/F interface. The latter did not exceed value of 1 nm showing high quality of the S/F interface. Using SQUID and mutual inductance setup we measured the superconducting transition temperatures for the samples, which are in agreement with the literature data for the given $d_S$ and $d_F$. Using PNR for the single S layer allowed us to obtain the screening length $\lambda$ of the superconducting layer, $\lambda = 120$nm. This value is 2.6 times higher than the London penetration depth for pure niobium. This can be explained by the mean free path $l$ of electrons in the S layer, $l \approx 6$nm, which is 7 times shorter than the BCS coherence length of clean niobium. Magnetic properties of the CuNi layers were studied by PNR and SQUID at $T > T_c$. The measurements have proved that all CuNi layers down to $d_F = 3.5$ nm are magnetic with the saturation magnetization $4\pi M = [0.2 \div 0.3]$ kGs.

Table 1. Structural, magnetic and superconducting parameters of the samples.

| No | $d_S$, nm | $d_F$, nm | $\sigma_{S/F}$, nm | $4\pi M_F$, kGs | $T_C$, K |
|---|---|---|---|---|---|
| 1 | 8.2 | 5.8 | 0.4 | 0.3 | N.A. |
| 2 | 8.0 | 3.5 | 0.8 | 0.3 | 5.4 |
| 3 | 8.3 | 6.9 | 0.2 | 0.2 | N.A. |
| 4 | 14.3 | 7.8 | 0.6 | 0.2 | 5.8 |
| 5 | 57 | 0 | - | 0 | 8.1 |

## 7. Acknowledgments

This work was carried out in the framework of the collaboration agreement between Max-Planck-Institut für Festkörperforschung and Institute of Electronic Engineering and Nanotechnologies ASM, and based upon experiments performed at the NREX instrument operated by Max-Planck Society at the Heinz Maier-Leibnitz Zentrum (MLZ), Garching, Germany. The neutron part of the project has been supported by the European Commission under the 7th Framework Programme through the "Research Infrastructures" action of the Capacities Programme, NMI3-II, Grant Agreement number 283883. R.M. and L.T. would like to thank the Program of Competitive Growth of Kazan Federal University among World's Leading Academic Centers. The SNMS depth profile measurements were supported by the Hungarian–Chinese bilateral project, TE'T_12_CN-1-2012-0036.